\documentstyle[amssymb,tighten,aps,epsf,floats]{revtex}

\begin{document}

\draft

\preprint{UTPT-97-04}

\title{The Initial-Value Problem of Spherically Symmetric Wyman Sector 
Nonsymmetric Gravitational Theory}

\author{M.\ A.\ Clayton,}
\address{CERN-Theory Division, CH-1211 Geneva 23, Switzerland}

\author{L.\ Demopoulos, and J.\ L\'egar\'e}
\address{Department of Physics, University of Toronto,
Toronto, Ontario, Canada M5S 1A7} 

\date{\today}

\maketitle

\begin{abstract}%
We cast the four-dimensional field equations of the Nonsymmetric
Gravitational Theory ({\sc ngt}) into a form appropriate for numerical
study.
In doing so, we have restricted ourselves to spherically symmetric
spacetimes, and we have kept only the
Wyman sector of the theory.
We investigate the well-posedness of the initial-value
problem of {\sc ngt} for a particular data set consisting of a 
pulse in the antisymmetric field on an asymptotically flat
space background.
We include some analytic results on the solvability of the 
initial-value problem
which allow us to place limits on the regions of the parameter
space where the initial-value problem is solvable.
These results are confirmed by numerically solving the constraints.
\end{abstract}


\section{Introduction}

Of late, there has been an increasing interest in numerical relativity 
and the application of numerical techniques to the study of 
dynamical gravitational objects.
The reason for this is
fairly clear: numerical relativity makes it possible to study 
physical situations that have defied analytic analysis.

The Nonsymmetric Gravitational Theory ({\sc ngt}) 
(see~\cite{bib:Moffat Phys Rev D 19 3554 1979,%
bib:Moffat Found Phys 14 1217 1984,%
bib:Moffat Banff,%
bib:Moffat Phys Lett B 355 447 1995,%
bib:Moffat J Math Phys 36 3722 1995,%
bib:Legare and Moffat Gen Rel Grav 27 761 1995}, 
as well as~\cite{bib:Moffat dynamical constraints} for the most
recent discussion)
is an alternate
theory of gravity that evolved from a re-interpretation of 
the Einstein Unified Field Theory 
({\sc uft}) 
(see~\cite{bib:Einstein Ann Math 46 578 1945,%
bib:Einstein and Straus Ann Math 47 731 1946,%
bib:Lichnerowicz J Rat Mech and Anal 3 487 1954,%
bib:Lichnerowicz 1955,%
bib:Maurer-Tison C R Acad Sc 242 1127 1956,%
bib:Hlavaty 1958,%
bib:Maurer-Tison Ann scient Ec Norm Sup 76 185 1959,%
bib:Tonnelat 1982}),
itself an extension of Einstein's General Relativity ({\sc gr}).
The theory proposes to do away with the assumption that the metric
tensor describing spacetime is a symmetric, second-rank tensor,
replacing it with a second-rank, nonsymmetric tensor.
This tensor is typically written as
the metric tensor in {\sc gr}: $g_{\mu\nu}$.
However, since a metric tensor is by definition a symmetric quantity,
$g_{\mu\nu}$ cannot represent the metric of the {\sc ngt} spacetime;
instead, we refer to it as the fundamental tensor 
(see the discussion 
in~\cite{bib:Lichnerowicz J Rat Mech and Anal 3 487 1954,%
bib:Lichnerowicz 1955,%
bib:Maurer-Tison C R Acad Sc 242 1127 1956}
and in 
particular~\cite{bib:Maurer-Tison Ann scient Ec Norm Sup 76 185 1959}).

Although on the surface the above-mentioned substitution appears
innocuous, in reality it induces fundamental changes in the underlying
structure of the theory, both physical and mathematical.
Indeed, because of the lack of a definite tensorial symmetry 
in the various quantities of the theory, the field equations of {\sc ngt}
are somewhat more complicated than the corresponding field equations
of {\sc gr}, as is the dynamical structure.
For instance, it can be shown that a spherically symmetric gravitational
field in {\sc gr} is
diffeomorphic to a static, spherically symmetric solution, so that all
spherically symmetric stellar objects have a static exterior.
This is known as Birkhoff's theorem (see for
instance~\cite{bib:Hawking and Ellis}, p.~372).
In contrast, it was demonstrated 
in~\cite{bib:Clayton J Math Phys 37 395 1996} that there is
no equivalent to Birkhoff's theorem in {\sc ngt}.
Hence, if we wish to study a stellar object even so simple as a 
spherically symmetric star in {\sc ngt}, we must consider the
full, time-dependent theory.

The increased mathematical complication of the field equations of
{\sc ngt} {\it versus} those of {\sc gr} has meant that {\sc ngt} has
been plagued since its inception by a distinct lack of exact, or in
some cases even approximate, solutions.
In particular, in addition to the (trivial) {\sc gr} solutions, 
a single static
solution to the spherically symmetric field equations
is known to exist for massless {\sc ngt}:
the so-called Wyman solution
(see~\cite{bib:Wyman Can J Math 2 427 1950,%
bib:Bonnor Proc Roy Soc 209 353 1951,%
bib:Vanstone Can J Math 14 568 1962,%
bib:Cornish and Moffat Phys Lett B 336 337 1994}).
The so-called Papapetrou solution
(see~\cite{bib:Papapetrou Proc Roy Soc Irish Acad Sci A 52 69 1948},
in particular~\S I) has been formally eliminated from the 
theory (see~\cite{bib:Moffat dynamical constraints}).
The addition of terms to the action by 
Moffat~\cite{bib:Moffat Phys Lett B 355 447 1995,%
bib:Moffat J Math Phys 36 3722 1995,%
bib:Legare and Moffat Gen Rel Grav 27 761 1995,%
bib:Moffat dynamical constraints}
does not simplify matters:
exact solutions for the massive theory are even more
elusive than for the massless theory.
It therefore appears that {\sc ngt} would be an ideal candidate for 
numerical investigations.

In this work, we begin the task of defining a formalism that will allow
the application of the techniques developed in the field of
numerical relativity to {\sc ngt};
the work was begun in~\cite{bib:Clayton thesis,%
bib:Clayton Int J Mod Phys D,%
bib:Clayton CQG 13 2851 1996},
where a Hamiltonian
formulation of {\sc ngt} was derived from the action principle
of the theory.
It was found in these references that the antisymmetric degrees of 
freedom separate into
two sectors, namely the Wyman sector and the Papapetrou sector.
One of these sectors,
specifically the Papapetrou sector, is plagued with
linearization instabilities;
this is, in fact, what motivated the work 
in~\cite{bib:Moffat dynamical constraints}, which eliminates 
these degrees of freedom.
However, since the two sectors of the theory 
are decoupled in evolution, by choosing Wyman sector
initial data, our results are equally applicable to most
{\sc uft}-like actions.

The next section gives a brief overview of the coordinate choices and
slicing conditions that allow us to rewrite the set of first-order field
equations in a form appropriate for numerical study.
In \S\ref{sec:solvability} we discuss the
initial-value problem, as well as some 
difficulties that 
arise when dealing with a constrained set of field equations, in
particular how those problems apply to {\sc ngt}.
In \S\ref{sec:Numerical_NGT} we discuss the numerical techniques relevant
to the solution of the initial-value problem of {\sc ngt}.
We also present one such generated solution and discuss its properties.
The first
appendix contains the spherically symmetric reduction of 
the field quantities, while the second appendix
gives the conformal transformation rules of the field equations.
The last appendix is a brief demonstration of two results used
in \S\ref{sec:solvability}.

\section{The Spherically Symmetric NGT Field Equations}

Early versions of 
{\sc ngt}~\cite{bib:Moffat Phys Rev D 19 3554 1979,%
bib:Moffat Found Phys 14 1217 1984,%
bib:Moffat Banff}
consisted of a re-interpretation of {\sc uft}, 
where the antisymmetric components of the fundamental tensor,
originally introduced to represent the components of the electromagnetic
field, were considered gravitational degrees of freedom.
The work in~\cite{bib:Damour Deser and McCarthy Phys Rev D 45 3289 1992,%
bib:Damour Deser and McCarthy Phys Rev D 47 1541 1993}
necessitated the addition of two terms to the action.
One of these, the so-called Bonnor term
(see~\cite{bib:Bonnor Proc R Soc A 226 366 1956}), endowed the 
antisymmetric fields with a mass and introduced a parameter
$\mu$ into the theory.
This parameter is 
referred to interchangeably as the {\sc ngt} mass
parameter or the inverse of the {\sc ngt} range parameter.
The findings of~\cite{bib:Clayton CQG 13 2851 1996} lead to a further
modification of the theory, described 
in~\cite{bib:Moffat dynamical constraints}.
This left $\mu$ as a free parameter of the theory, with a 
potential phenomenological interpretation
in applications such as galaxy
dynamics (see~\cite{bib:Moffat and Sokolov Phys Lett B 378 59 1996}).

Rather than re-iterating the full set of four dimensional
{\sc ngt} field equations here, 
we will simply refer the interested reader to the literature
(see, for instance,~\cite{bib:Moffat Phys Rev D 19 3554 1979,%
bib:Moffat Found Phys 14 1217 1984,%
bib:Moffat Banff,%
bib:Moffat Phys Lett B 355 447 1995,%
bib:Moffat J Math Phys 36 3722 1995,%
bib:Legare and Moffat Gen Rel Grav 27 761 1995,%
bib:Moffat dynamical constraints},
\cite{bib:Clayton J Math Phys 37 395 1996}, and
\cite{bib:Clayton thesis}, among others).
Instead, we offer in Appendices~\ref{sec:Appendix_First-Order_form}
and~\ref{sec:Appendix_Conformally-transformed_equations}
an overview of the spherically symmetric, Wyman sector
field equations written in
first-order form;
these will be sufficient for our purposes.
These appendices serve also to define the notation we use in this work.
The goal of the current section 
is to make the coordinate choices and to choose 
the slicing conditions that will reduce the {\sc ngt} field equations to
a form amenable to numerical study.

Upon performing a conformal transformation of the field variables:
$\gamma^{ab} \rightarrow \phi^{-4}\gamma^{ab}$ and
$K_{ab} \rightarrow \phi^{-2}K_{ab}$, 
where $\gamma^{ab}$ is the inverse of the hypersurface metric and
$K_{ab}$ is its extrinsic curvature 
(for details, see 
Appendix~\ref{sec:Appendix_Conformally-transformed_equations}),
we find that, just as in {\sc gr}, the Hamiltonian and momentum constraints
remain inter-twined.
In order to make any headway in solving these constraints,
we must find a way of decoupling them.
In {\sc gr}, this is accomplished, for example, by using maximal slicing:
${\rm Tr}[K] = K^\mu{}_\nu = \gamma^{\mu\nu} K_{\mu\nu} = 0$, which
fully decouples the Hamiltonian and momentum constraints.
Preserving this condition in time, $\partial_t[{\rm Tr}[K]] = 0$, yields
a second-order elliptic equation for the lapse function:
$\gamma^{ab}\nabla_a \nabla_b[N] - NR^{(3)} = 0$ (see, for 
instance,~(3.4) in \S III.B 
of~\cite{bib:Smarr and York Phys Rev D 17 2529 1978}).
A similar approach can be taken in {\sc ngt}; however, the resulting
``lapse equation'' is not nearly as simple as its {\sc gr} counterpart.
Instead, it turns out to be more profitable to 
take\footnote{From Appendix~\ref{sec:Appendix_First-Order_form}, 
$K^2{}_2 = \gamma^{22}K_{22} + \gamma^{[23]} j_{[23]}$
and $j^2{}_3 = \gamma^{[23]} K_{22} - \gamma^{22} j_{[23]}$.}
$K^2{}_2 = 0$.
In this case, the symmetric-sector components of the extrinsic curvature
drop out completely from the Hamiltonian constraint.
Although both the conformal factor and the extrinsic curvature still
appear in
the momentum constraint, it is only the $K_{11}$ component of the latter
that appears.
Therefore, if a solution to the Hamiltonian constraint can be found,
this can be inserted into the momentum constraint, which can then be solved
for the remaining component of the extrinsic curvature.

In spherical symmetry, there is only one coordinate choice to be made:
a radial coordinate $r$ is chosen such that
$[(\gamma^{22})^2 + (\gamma^{[23]})^2]^{1/2} = R^{-2}(r)$,
where $R(r)$ is some currently unspecified function;
we will choose $R(r) = r$.
This choice can obviously be satisfied by taking
$\gamma^{22} = R^{-2}(r)\cos\psi$ and $\gamma^{[23]} = R^{-2}(r)\sin\psi$,
or $\tan\psi = \gamma^{[23]} / \gamma^{22}$.
The components of the surface Ricci tensor are then
(see Appendix~\ref{sec:Appendix_First-Order_form})
\begin{eqnarray*}
R^{{\rm NS}(3)}_{11}
&=& -2\partial^2_r[\ln R] - 2(\partial_r[\ln R])^2
- \case{1}{2}(\partial_r[\psi])^2 
- \partial_r[\ln \gamma^{11}]\partial_r[\ln R] , \\
R^{{\rm NS}(3)}_{22}
&=& 1 - R^2\gamma^{11}(\cos\psi\partial_r^2[\ln R]
- 2 \cos\psi(\partial_r[\ln R])^2 - \case{1}{2}\sin\psi\partial_r^2[\psi]
- \sin\psi\partial_r[\ln R]\partial_r[\psi]) \nonumber \\
& & \qquad\hbox{} 
- \case{1}{2} R^2\partial_r[\gamma^{11}]
(\cos\psi \partial_r[\ln R] + \case{1}{2}\sin\psi \partial_r[\psi]) , \\
\noalign{\noindent and}
R^{{\rm NS}(3)}_{[23]}
&=& -R^2\gamma^{11}(\sin\psi \partial_r^2[\ln R]
- 2 \sin\psi (\partial_r[\ln R])^2 + \case{1}{2}\cos\psi \partial_r^2[\psi]
+ \cos\psi \partial_r[\ln R]\partial_r[\psi]) \\
& & \qquad\hbox{} 
- \case{1}{2} R^2 \partial_r[\gamma^{11}]
(\sin\psi \partial_r[\ln R] - \case{1}{2}\cos\psi \partial_r[\psi]) , 
\end{eqnarray*}
from which the Ricci scalar is found to be
\begin{equation}
\label{eq:Ricci_scalar}
R^{{\rm NS}(3)}
= -2\gamma^{11}(2\partial_r^2[\ln R] + 3 (\partial_r[\ln R])^2)
+ 2 R^{-2}\cos\psi - \case{1}{2}\gamma^{11}(\partial_r[\psi])^2
- 2\partial_r[\gamma^{11}]\partial_r[\ln R] .
\end{equation}
We also define
\begin{eqnarray*}
R^2{}_2
&=& \gamma^{22} R^{{\rm NS}(3)}_{22} + \gamma^{[23]} R^{{\rm NS}(3)}_{[23]}
= R^{-2}\cos\psi - \gamma^{11}\partial_r^2[\ln R]
- 2 \gamma^{11} (\partial_r[\ln R])^2
- \case{1}{2}\partial_r[\gamma^{11}]\partial_r[\ln R] , \\
\noalign{\noindent and}
R^2{}_3
&=& \gamma^{[23]} R^{{\rm NS}(3)}_{22} - \gamma^{22} R^{{\rm NS}(3)}_{[23]}
= R^{-2}\sin\psi - \case{1}{2}\gamma^{11}\partial_r^2[\psi]
- \gamma^{11}\partial_r[\ln R]\partial_r[\psi]
- \case{1}{4}\partial_r[\gamma^{11}]\partial_r[\psi] .
\end{eqnarray*}

As mentioned above,
if we choose the slicing condition $K^2{}_2 = 0$, the Hamiltonian and
momentum constraints are effectively
decoupled, allowing them to be solved 
independently.\footnote{Strictly speaking, the Hamiltonian constraint 
must be solved first, and its solution must be used to solve the momentum
constraint.}
In fact, 
using the definition of 
the derivative operator $\Delta_\gamma[\,]$ given
in~(\ref{eq:derivative_definition}),
the Hamiltonian constraint is written
\begin{mathletters}
\label{eq:Hamiltonian_constraint_both_forms}
\begin{eqnarray}
\label{eq:Hamiltonian_constraint_Delta}
{\cal H} = 0 
&=& [\gamma^{11}]^{-1/2}\phi
[8\Delta_\gamma[\phi] - \phi R^{{\rm NS}(3)}
+ 2 R^2\phi^{-7}[(j^2{}_3)^2 - K^2{}_2(K^2{}_2 + 2 K^1{}_1)]
+ \case{1}{4}\mu^2 R^2\phi^5\sin^2\psi] \\
&=& [\gamma^{11}]^{-1/2}\phi
[8R^2\gamma^{11}\partial_r^2[\phi] 
+ 16R\gamma^{11}\partial_r[\phi]
+ 4R^2\partial_r[\phi]\partial_r[\gamma^{11}]
+ \case{1}{2}R^2\phi\gamma^{11}(\partial_r[\psi])^2 \nonumber \\
& & \label{eq:Hamiltonian_constraint}
\qquad\hbox{}
+ 2R\phi\partial_r[\gamma^{11}] - 2\phi\cos\psi
+ 2\phi\gamma^{11} + 2R^2(j^2{}_3)^2\phi^{-7}
+ \case{1}{4}\mu^2 R^2\phi^5\sin^2\psi] ,
\end{eqnarray}
\end{mathletters}%
while the momentum constraint becomes
${\cal H}_1 
= -2R^2[\gamma^{11}]^{-1/2}(\gamma^{11}K_{11}\partial_r[\ln(R^2\phi^4)]
- j^2{}_3\partial_r[\psi]) = 0$.
The latter one of these is solved for the 
sole remaining symmetric-sector
component of the extrinsic curvature in a straightforward manner:
\begin{equation}
\label{eq:solved_momentum_constraint}
K_{11} = \frac{j^2{}_3 \partial_r[\psi]}{\gamma^{11} 
\partial_r[\ln(R^2\phi^4)]} .
\end{equation}
That the momentum constraint can be solved for $K_{11}$ in
this fashion allows us to implement a ``semi-constrained''
evolution.
What is more, 
by setting $\phi = 1$ in~(\ref{eq:Hamiltonian_constraint}), we
obtain a first-order, ordinary differential equation for $\gamma^{11}$, 
whose solution on every time-slice would generate the evolution of
that field variable.
This would allow us to implement a fully constrained evolution,
eliminating both the Hamiltonian and momentum constraints,
as well as the symmetric-sector evolution equations,
hence by-passing the problems inherent in evolving a 
set of discretized field equations (see, for instance, the discussion
of the Cauchy problem in~\cite{bib:Stewart 1984}, as well as the
issues raised 
in~\cite{bib:Bernstein Hobill and Smarr 1989}
and~\cite{bib:Anninos et al Phys Rev D 51 5562 1995} on the
maintenance of constraints during numerical evolution).
However, at this time
there does not seem to be any obvious advantage in doing this, 
and it would only
further cloud the issue of the solvability of the 
Hamiltonian constraint (see~\S\ref{sec:solvability} below).
In this work, we have chosen to implement the semi-constrained
evolution mentioned above.

Naturally, the slicing condition $K^2{}_2 = 0$ and the
coordinate choice 
$[(\gamma^{22})^2 + (\gamma^{[23]})^2]^{1/2} = R^{-2}(r)$
mentioned above
must be enforced:
$\partial_t[K^2{}_2] = 0$ 
and~$\partial_t[R] = 0$.
The first of these leads to an equation for the lapse function:
\begin{equation}
\label{eq:lapse_equation}
\frac{\partial_r[N]}{N}
= -\frac{R}{\gamma^{11} \partial_r[R\phi^2]}
[\gamma^{11}\partial_r^2[\phi^2] 
+ R^{-6}\phi \partial_r[\phi]\partial_r[\gamma^{11}R^6]
+ \case{1}{4} \mu^2 \phi^6 \sin^2\psi] .
\end{equation}
This first-order, ordinary differential equation is solved on every
time-slice for the lapse function.
Meanwhile, preserving the coordinate condition in time
leads to an equation for the only non-trivial component of
the shift function:
\begin{equation}
\label{eq:shift_equation}
\partial_r[N^1] - 2 N^1 \partial_r[\ln(R\phi^2)]
- 2 N \phi^{-6} K^2{}_2 = 0 .
\end{equation}
When $K^2{}_2 = 0$, 
the solution of this is $N^1 = AR^2\phi^4$, where $A(t)$ is a constant
of integration;
here, we take $A(t) = 0$, so that $N^1 \rightarrow 0$ as 
$r\rightarrow +\infty$, maintaining asymptotically flat spatial slices
in evolution.

Once the slicing conditions, coordinate choices, and constraints
are taken care of, the evolution equations are:
\begin{mathletters}
\label{eq:evolution_equations}
\begin{eqnarray}
\partial_t[\gamma^{11}]
&=& -2N\phi^{-6}[\gamma^{11}]^2 K_{11} \\
\partial_t[\psi]
&=& 2N\phi^{-6}j^2{}_3 \\
\partial_t[K_{11}]
&=& 4N[\phi\partial_r^2[\phi] - (\partial_r[\phi])^2]
+ 2\phi\partial_r[\phi](N\partial_r[\ln\gamma^{11}]
+ 2N\partial_n[\ln R] - \partial_r[N]) 
+ \phi^2\partial_r^2[N] \nonumber \\
& & \qquad\hbox{} 
+ \case{1}{2}\phi^2\partial_r[N]\partial_r[\ln\gamma^{11}]
- N\phi^2 R^{{\rm NS}(3)}_{11} + N\phi^{-6}\gamma^{11}[K_{11}]^2 
+ \case{1}{4}\mu^2N\phi^6[\gamma^{11}]^{-1} \sin^2\psi \\
\partial_t[j^2{}_3]
&=& N\phi\gamma^{11}\partial_r[\phi]\partial_r[\psi]
+ \case{1}{2}\phi^2\gamma^{11}\partial_r[N]\partial_r[\psi]
- N\phi^2 R^2{}_3
- N\phi^{-6}\gamma^{11} K_{11} j^2{}_3
- \case{1}{2}\mu^2 N \phi^6 \sin\psi \cos\psi .
\end{eqnarray}
\end{mathletters}%
Note that we have included here the evolution equation for 
$K_{11}$, for the sake of completeness.

\section{The Initial-Value Problem in NGT 
for Spherically Symmetric Systems}
\label{sec:solvability}

In the words of York and Piran,
``The initial value problem of a physical theory is ascertaining what 
data must be specified at a given time in order that the equations of
motion determine uniquely the evolution of the system.'' 
(see~\cite{bib:York and Piran}, p.~147).
In formulating {\sc ngt} as a Hamiltonian system, Clayton
(see~\cite{bib:Clayton thesis}) laid most of the groundwork for
the study of its initial-value 
problem.
As with most Lagrangian field theories, the field equations of 
{\sc ngt} separate into a set of initial-value constraints
that impose diffeomorphism invariance, and a set of evolution equations.
In this section, we are concerned with the initial-value constraints
of {\sc ngt},
also known as the Hamiltonian and momentum constraints.

It was demonstrated in the previous section that the momentum constraint 
can be explicitly satisfied, as in~(\ref{eq:solved_momentum_constraint}),
and we need not be concerned with it any further.
Meanwhile, the Hamiltonian constraint is a second-order, ordinary
differential equation for $\phi$, the conformal factor.
Unlike the momentum constraint, the solvability of the Hamiltonian
constraint is by no means guaranteed.
Indeed, we can gain some insight by recognizing, as did
Wheeler (see~\cite{bib:Wheeler 1964,%
bib:Murchadha and York Phys Rev D 10 2345 1974})
the similarity between the Hamiltonian constraint (for a moment of
time-symmetry)
and a scattering problem in
non-relativistic Schr\"odinger mechanics,
with the curvature scalar playing the role of the potential.
If $R^{{\rm NS}(3)}$ is positive everywhere, then the Hamiltonian will
always have a solution, corresponding to scattering off a potential
barrier.
If the curvature scalar is negative but ``shallow'' enough, 
then a scattering solution will also exist.
Only if $R^{{\rm NS}(3)}$ is negative and ``deep'' enough will
a bound state begin to form, and hence no scattering solution will exist.
This is the type of behaviour that will be manifest itself below:
there are actually situations where the antisymmetric fields
could be considered
perturbations on a {\sc gr} background, 
in the sense that they are by 
no means large, in which the Hamiltonian constraint has no
asymptotically flat solution. 
Of course, the reason for this behaviour is that the derivatives
of the antisymmetric variables enter the curvature scalar, along with
the variables themselves.
Thus, one can choose values for the antisymmetric variables that are
small in overall magnitude, but that are sharply peaked, causing a
corresponding ``deepening of the well'',
which in turn causes the 
Hamiltonian constraint to become unsolvable for an asymptotically
flat spacetime.\footnote{In such a case, a solution would most likely
exist for a non-asymptotically flat manifold or a closed 
manifold (as in~\cite{bib:Murchadha CQG 4 1609 1987});
we are not considering that option here.}

In Appendix~\ref{sec:Appendix_proof_of_criterion}, 
we derive two results that
are consequences of 
the existence of a solution
to the Hamiltonian constraint, given the
appropriate boundary conditions.
In other words, the solvability of the Hamiltonian constraint implies
that the inequalities~(\ref{eq:strong_condition_on_phi})
and~(\ref{eq:weak_condition_on_phi}) are satisfied.
However, it is important to keep in mind that these inequalities hold
only for massless {\sc ngt}, when $\mu = 0$.
The mass term in~(\ref{eq:Hamiltonian_constraint_both_forms}) 
makes it considerably
more difficult to generate similar inequalities for the case of
massive {\sc ngt}.
Furthermore, we emphasize that these inequalities are consequences of
the solvability of the Hamiltonian constraint:
to demonstrate that these inequalities actually imply 
that a solution to
the Hamiltonian constraint exists is a somewhat more involved process,
and will be considered in the future.

Given an initial data set, $\{\gamma^{11}, \psi, j^2{}_3\}$,
we can use~(\ref{eq:strong_condition_on_phi}) to place a limit on the
field variables that will allow us to solve
the Hamiltonian constraint for an asymptotically flat spacetime.
The procedure is actually one of elimination: if we find values of the
field variables for which~(\ref{eq:strong_condition_on_phi}) is not
verified, then we may conclude that the Hamiltonian constraint cannot
be solved for such an initial data set, as the proof 
in Appendix~\ref{sec:Appendix_proof_of_criterion}
concludes that if
the Hamiltonian constraint is solvable, 
then~(\ref{eq:strong_condition_on_phi}) is 
verified.\footnote{It 
is the converse which is not necessarily 
true:~(\ref{eq:strong_condition_on_phi}) might very well be 
verified, yet the Hamiltonian constraint might not possess a solution.}
This process is continued, making the initial data set evermore 
conservative, for instance by moving closer and closer to a Minkowski
space initial data set, 
until~(\ref{eq:strong_condition_on_phi}) is verified.
At this point, we cannot conclude that a solution of the Hamiltonian
constraint exists: rather, we conclude that a solution certainly
{\em does not} exist for less conservative initial data.

If this procedure is to help us in placing limits on the solvability of
the Hamiltonian constraint, then we require a 
(non-trivial) function $f$ that
makes the left-hand side of~(\ref{eq:strong_condition_on_phi}) a
minimum.
This function is 
well-known~\cite{bib:O Murchadha PC}:
$f(r) = [\xi / (\xi^2 + r^2)]^{1/2}$.
Here $\xi$ is a constant, and the factor of $\xi^{1/2}$ in the 
numerator is included to make the value of the left-hand side 
of~(\ref{eq:strong_condition_on_phi}) independent of 
$\xi$;~(\ref{eq:strong_condition_on_phi}) is obviously invariant under
such a scaling of $f$.
The value of $\xi$ is then used to make the value of the 
right-hand side of~(\ref{eq:strong_condition_on_phi}) as large as possible.
Note in particular that the maximum of $f$ occurs at $\xi = \pm r$.

As noted in Appendix~\ref{sec:Appendix_proof_of_criterion}, 
the two inequalities~(\ref{eq:strong_condition_on_phi}) 
and~(\ref{eq:weak_condition_on_phi}) are used to different ends.
We mentioned above that~(\ref{eq:strong_condition_on_phi})
is used to find a {\em most} conservative initial data set
$\{\gamma^{11}, \psi, j^2{}_3\}$ for which the Hamiltonian constraint
is guaranteed to be unsolvable.
On the other hand,~(\ref{eq:weak_condition_on_phi}) is used to find
a {\em least} conservative initial data set for which the Hamiltonian
constraint is guaranteed to be solvable.
In a sense, the two inequalities,~(\ref{eq:strong_condition_on_phi})
and~(\ref{eq:weak_condition_on_phi}) are used to ``sandwich'' the 
limiting initial data set where the Hamiltonian constraint is just
barely solvable.

In this work, we are considering initial data sets in which
$\gamma^{11} = 1$, $j^2{}_3 = 0$, and where $\psi$ has the form of a 
pulse:
\begin{equation}
\label{eq:not_so_Gaussian_pulse}
\psi(r) = A [r / r_0]^2 e^{-(r - r_0)^2 / \sigma^2} .
\end{equation}
The dimensionless parameter $A$ gives the overall amplitude
of the pulse, while $r_0$ and $\sigma$ are lengths that fix the 
position and the width, respectively, of the pulse;
the location of the maximum of the pulse is
$r_{\rm peak} = \case{1}{2}[r_0 + [r_0^2 + 4\sigma^2]^{1/2}]$.
This particular function was chosen because the Gaussian factor
causes both $\psi$ and its derivatives to vanish quite rapidly away from
$r_0$, while the factor of $[r/r_0]^2$ ensures that the Ricci curvature
scalar in~(\ref{eq:Ricci_scalar}) remains finite as $r\rightarrow 0$.
In a sense, the pulse is ``semi-localized'' about $r=r_{\rm peak}$, 
allowing us to study the properties of {\sc ngt} initial data without
having to worry about this will affect the boundary conditions.
For such an initial data set, the curvature scalar is more or less
peaked about $r = r_0$.
We therefore choose $\xi = r_{\rm peak}$, 
in order to maximize the value of the integrand,
$f^2R^{{\rm NS}(3)}$.
The remaining three free parameters, $A$, $\sigma$, and $r_0$,
are adjusted in the manner discussed above until the inequality 
in~(\ref{eq:strong_condition_on_phi}) is no longer verified.
For instance, since taking $A = 0$ reduces the initial data set to
that of Minkowski space, then we expect to find some maximum value 
$A = A_{\rm max}$ which renders the Hamiltonian constraint unsolvable.
Of course, this is intuitively obvious: if there is to be a data set
for which the Hamiltonian constraint cannot be solved for an asymptotically
flat spacetime, then we expect that data set to correspond to a strong-field
region.
However, as was mentioned above, it is not only the overall scale of the
antisymmetric variables that affects the solvability of the Hamiltonian
constraint, but also the scale of their derivatives.
In particular, smaller values of the width parameter $\sigma$ will cause
$\psi$ to become correspondingly more peaked, and the ensuing increase
in its derivative could feasibly make the Hamiltonian constraint 
unsolvable.

As we approach a region of unsolvability in the parameter space, the
mass of the system, as measured by an observer in the asymptotically
flat region, begins to increase without bound.
This is to be expected: a large value of the mass would arise out of a 
spacetime of such great curvature that it can no longer be represented by
an asymptotically flat manifold.
It is therefore of great interest to measure the mass of the system as a
function of the parameters of the pulse defining $\psi$.
Through an analysis of the surface terms that arise out of the canonical
decomposition of the action~\cite{bib:Clayton thesis}, it can be shown
that the
{\sc adm} mass (see~\cite{bib:Arnowitt Deser and Misner}) 
for asymptotically flat initial data\footnote{For our purposes, 
asymptotically flat initial data is defined by the usual conditions on
the symmetric-sector functions (as in {\sc gr}), and ``faster'' fall-off 
in the antisymmetric-sector variables.
This is defined explicitly in~\cite{bib:Clayton thesis}, p.~116.}
is
\begin{equation}
\label{eq:ADM_mass_in_NGT_in_general}
M = \bar M - \frac{1}{2\pi}\oint_{\partial\Sigma} \gamma^{ab}
\nabla_a [\phi]
\, dS_b
\end{equation}
where $\partial\Sigma$ is the boundary of the spatial slice and where
$\bar M$ is the {\sc adm} mass of the conformal
metric and $M$ is the {\sc adm} mass of the physical metric.
For the conformally flat field variables we consider, and given the 
fall-off that we have assumed, it is a simple matter to demonstrate 
that~(\ref{eq:ADM_mass_in_NGT_in_general}) reduces to
\begin{equation}
\label{eq:ADM_mass_in_NGT}
M = -2\int_0^{+\infty} \Delta_\gamma[\phi] \, r^2 dr
\end{equation}
where the derivative operator $\Delta_\gamma[\,]$ is defined 
in~(\ref{eq:derivative_definition}).
Once a solution to the Hamiltonian constraint is found, it can be
inserted into~(\ref{eq:ADM_mass_in_NGT}), and the mass $M$ calculated;
note that this mass is guaranteed to be positive 
as $R^{{\rm NS}(3)}$ 
and the Bonnor term are negative definite.
Alternately, if we assume that $\phi \rightarrow 1 + M / 2r$ asymptotically,
where $M$ is a constant, we can read off the value of $M$ by inspecting
the behaviour of $\phi$ in the asymptotic region.

In the next section, we discuss a numerical scheme that can be used to
analyze the initial-value problem of {\sc ngt}.
In addition to presenting a typical solution to the
Hamiltonian constraint, we use this scheme and the 
conditions discussed above to place limits on the solvability of the
initial-value problem in {\sc ngt}.

\section{Numerical Analysis of the Initial-Value Problem in NGT}
\label{sec:Numerical_NGT}

As written, the Hamiltonian constraint~(\ref{eq:Hamiltonian_constraint})
is a second-order, ordinary differential equation for the conformal
factor $\phi$.
There are two important points to note. 
Firstly, suppose we choose
{\sc gr} initial conditions: {\it i.e.}, $\gamma^{11} = 1$, 
$R = r$ (standard spherical coordinates), and
vanishing antisymmetric-sector data ($\psi = 0$ and $j^2{}_3 = 0$).
The Hamiltonian constraint then reduces to
$r^2\partial_r^2[\phi] + 2 r \partial_r[\phi] 
= \partial_r[r^2 \partial_r[\phi]] = 0$,
whose solution is $\phi = 1 + M / 2r$, where $M$ is a constant of 
integration, physically interpreted as being the 
mass defined by~(\ref{eq:ADM_mass_in_NGT}).
Secondly, when $\mu \ne 0$, the Hamiltonian constraint is a non-linear,
second-order,
ordinary differential equation for the conformal factor, in contrast to
{\sc gr}, where the Hamiltonian constraint for a vacuum spacetime
is linear in the conformal factor when the extrinsic curvature vanishes
(moment of time symmetry); to see the non-linear behaviour,
matter or energy of some kind must be added to the system.
Of course, this non-linearity in the Hamiltonian constraint 
is a manifestation of the fact that, loosely speaking,
the antisymmetric-sector fields behave like matter or energy to the 
symmetric-sector fields, albeit with a nontrivial interaction.
Fortunately, since the Hamiltonian constraint in
spherical symmetry is an ordinary differential
equation, we can use a fairly simple scheme for its solution, and it is
this procedure that we describe in this section.

Since we are considering initial data that are concentrated in an
area of the initial slice, in those areas of the initial slice where
the initial data of the antisymmetric sector are vanishing or nearly so, 
$\phi$ will be more or less approximated by its {\sc gr}
solution, $\phi \approx 1 + M / 2r$, where $M$ is either some constant
or a linear function of $r$.
The first case corresponds to a Schwarzschild-like region of the initial
slice, while the second case is physically equivalent to a flat, Minkowski
region of the initial 
slice.\footnote{If $M$ is a linear 
function of $r$, then $M/2r$ is some constant,
and therefore $\phi$ is a constant other than unity.
However, by rescaling the basis vectors, we can absorb this constant and
make $\phi = 1$, just as in the case of Minkowski space.}
We would expect to find the former behaviour in the asymptotic region, 
while the latter behaviour would be prominent near the origin, given
our choice of $\psi$.
The only region which remains in question is the intermediate region,
where the antisymmetric-sector initial data are non-trivial;
however, this essentially becomes a region of transition that serves
the purpose of matching the Minkowski-type solution near the origin to
the Schwarzschild-type solution of the exterior.
The obvious generalization is 
therefore to assume that $M$ no longer has this simple behaviour,
and to make the substitution $M \rightarrow M(r)$. 
We will therefore use $\phi(r) = 1 + M(r) / 2r$ for all $r$, and solve
the Hamiltonian constraint for the function $M(r)$.

Upon substituting $\phi = 1 + M(r) / 2r$ into the Hamiltonian 
constraint~(\ref{eq:Hamiltonian_constraint}), we obtain 
\begin{equation}
\label{eq:M_equation}
\partial_r^2[M]
= \frac{\partial_r[\ln\gamma^{11}]}{2r} (M - r\partial_r[M])
+ \frac{1}{2}\left[
\frac{1}{r}\left(\frac{\cos\psi}{\gamma^{11}} - 1\right)
- \partial_r[\ln\gamma^{11}] - \case{1}{4}r(\partial_r[\psi])^2\right]\phi
- \frac{r (j^2{}_3)^2}{2\gamma^{11}\phi^7}
- \frac{\mu^2 r\phi^5\sin^2\psi}{16\gamma^{11}} ,
\end{equation}
where $\phi$ is to be considered shorthand for $\phi = 1 + M(r) / 2r$.
This equation is finite-differenced across the spatial grid, using
centred differences for both the first- and second-order derivatives,
\[
\partial_r[M] 
\rightarrow \frac{\delta(\mu M)_i}{\delta(\mu r)_i}
= \frac{M_{i+1} - M_{i-1}}{2\Delta r} 
\qquad\hbox{\rm and}\qquad
\partial_r^2[M] 
\rightarrow \frac{\delta}{\delta r_i}
\left[\frac{\delta M_i}{\delta r_i}\right]_i
= \frac{M_{i+1} - 2 M_i + M_{i-1}}{\Delta r} ,
\]
where $\Delta r$ is the (uniform) spatial grid 
spacing.\footnote{The notation is standard: see, for instance,
\cite{bib:Celia and Gray}, p.~48,
for definitions of the centred-differencing
operator, $\delta$, and the averaging operator, $\mu$, as well as the
forward- and backward-differencing operators,
$\vartriangle$ and $\triangledown$, which will be used later.}
This
yields an equation of the form
$M_{i+1} - 2M_{i} + M_{i-1} = f(M)$, where the notation $f(M)$ is
used to represent the fact that the right-hand side has some 
functional dependence on
the value of $M$ at grid points $i+1$, $i$, and $i-1$, along with the
other variables.
An implicit relaxation scheme is used to solve this 
equation\footnote{Such schemes are described in the standard 
references; see, for instance, \cite{bib:Fox}, p.~86.}:
beginning with an initial guess $M_i^0$ for the $M_i$, we write
$M_{i+1}^j - 2M_i^j + M_{i-1}^j = f(M^{j-1})$ and solve for 
the $M^j$.
This process is repeated until a precision criterion is reached:
in our case, we chose to calculate the integral of the left and right sides
of~(\ref{eq:M_equation}) for a given iteration.
If the difference between these two is smaller than some predetermined
value, then the solution has converged and the iteration ceases.
The accuracy of the results will be discussed at the end of this
section.

Evidently,~(\ref{eq:M_equation}) and its finite-differenced counterpart
are to be augmented by boundary conditions on $M$ that will fix the 
exact physical situation that is being studied.
In this work, we have chosen to take $M$ to be linear in $r$ near the origin.
Numerically, this boundary condition is most easily implemented by noting
that if $M$ is linear in $r$, then the average value of $M$ 
at a given point coincides with the value of $M$ itself at that
point: $(\mu M)_i = M_i$.
Evaluating this at the origin, where $M = 0$, we have
$(\mu M)_0 = 0$, where we have labeled the origin by gridpoint $i=0$.
In the asymptotic region, we have taken $M$ to behave like a constant,
and thus its derivative must vanish: $\partial_r[M] = 0$.
This guarantees that $\phi \rightarrow 1$ as $r \rightarrow +\infty$.
Numerically, this is encoded using either a backward difference
based on the grid point $n$, or a forward difference based on the
grid point $n - 1$, where $n$ labels the outer-most grid point: because
the derivative is made to vanish, both formulations are equivalent.
Therefore, in the asymptotic region, we take either
$(\triangledown M)_n = 0$ or $(\vartriangle\! M)_{n-1} = 0$.

\begin{figure}
\centerline{\epsffile{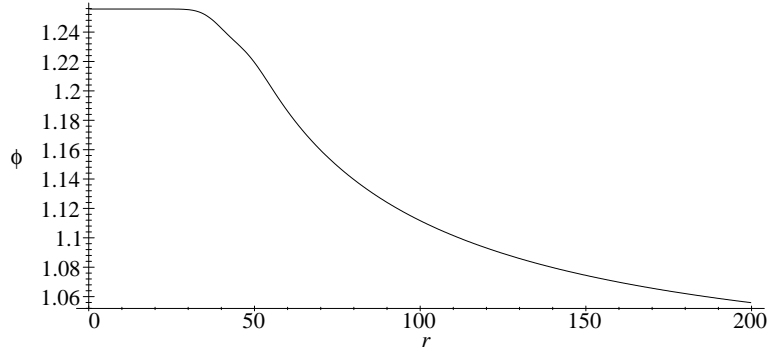}}
\caption{The conformal factor $\phi$ of massless {\sc ngt}
as a function of radial distance
from the origin, for the initial conditions
$\gamma^{11} = 1$, $\psi = A [r/r_0]^2 e^{-(r - r_0)^2 / \sigma^2}$, 
and~$j^2{}_3 = 0$.
We have taken $A = 0.7$ and
$r_0 = 40$ in units where $\sigma = 10$.
The radial variable $r$ is measured in the same units as $r_0$.
These data have an {\sc adm} mass of $M \approx 22.340$ in the
same units as $r$.}
\label{fig:conformal_factor_sample}
\end{figure}

\begin{figure}
\centerline{\epsffile{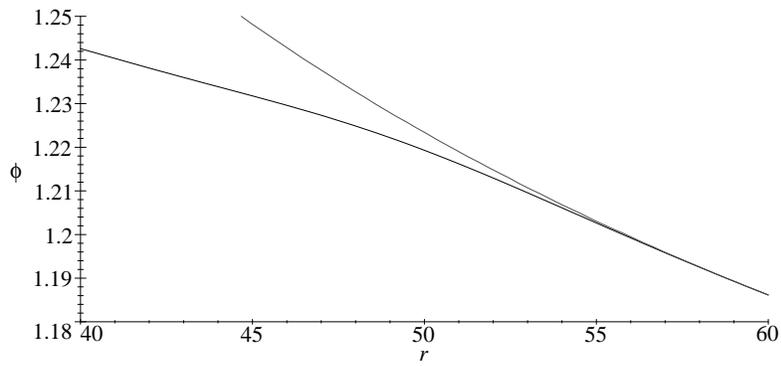}}
\caption{For the same initial data and parameters as
Figure~\protect\ref{fig:conformal_factor_sample},
we overlay a magnified region of that graph (lower curve) onto
a plot of the {\sc gr} conformal factor, $1 + M / 2r$, where
$M \approx 22.340$ in the same units as $r$.}
\label{fig:conformal_factor_sample_blow-up}
\end{figure}

In Figure~\ref{fig:conformal_factor_sample}, we give a plot of the
conformal factor of massless {\sc ngt} for the initial conditions
$\gamma^{11} = 1$ and $j^2{}_3 = 0$, with $\psi$ given 
by~(\ref{eq:not_so_Gaussian_pulse}) with $A = 0.7$ and
$r_0 = 40$ in units where $\sigma = 10$; 
in this and in every other case in this paper, the units of $r$,
$r_0$, $\sigma$, and $M$ will be assumed to be the same.
The solution was found using the implicit relaxation scheme described
above.
The values of the parameters were chosen to place the simulation in
an intermediate region of the parameter space: sufficiently far away
from Minkowski space so as to avoid making the simulation trivial, and
sufficiently far away from the regions of unsolvability discussed
in \S\ref{sec:solvability} to avoid infringing on some limiting case
that could raise questions about the stability of the numerical solution.
Throughout, we have chosen $r_{\rm max}$, the location of the outermost
grid point, to be at least $r_{\rm peak} + 10\sigma$: this guarantees
that $\psi(r_{\rm max}) \approx 0$ and that we can impose asymptotically
flat boundary conditions.
The grid spacing was taken to be $\Delta r = 0.015$ in the same units
as $r$.
In other words, the solution depicted in 
Figure~\ref{fig:conformal_factor_sample}
represents a generic conformal factor for the initial data set used in
this work, with the antisymmetric sector variables being strong, but not
pathologically so.
The {\sc adm} mass of the system (see the discussion in
\S\ref{sec:solvability}, in particular~(\ref{eq:ADM_mass_in_NGT_in_general})
and~(\ref{eq:ADM_mass_in_NGT}) for the definition of the
{\sc adm} mass in {\sc ngt})
is calculated to be
$M \approx 22.340$.
The graph of Figure~\ref{fig:conformal_factor_sample_blow-up} overlays
the {\sc gr} conformal factor $1 + M / 2r$, where $M \approx 22.340$, 
atop a magnified section of the conformal factor generated from the
numerical code.
This graph demonstrates that $M(r)$ goes to a constant in the asymptotic
region, despite the large width ($\sigma = 10$) of the initial data.

The solution displayed in Figure~\ref{fig:conformal_factor_sample} 
evidently possesses
many of the features described earlier.
In fact, the radial axis is clearly divided into three regions, corresponding
to the central peak of $\psi$, and the regions a few widths $\sigma$
above and below this peak.
For our choice of parameters, the central peak of $\psi$ is located at
$r_{\rm peak} \approx 42.36$.
The region $0 < r \lesssim 30$ corresponds to a Minkowski space solution,
while the region $r \gtrsim 60$ corresponds to a Schwarzschild-type solution,
$\phi = 1 + M/2r$, where $M$ is a constant; in this case,
$M \approx 22.34$.
In the latter region, which is at least two widths away from the
central peak at $r_{\rm peak}$, the value of $\psi$ is, by
construction, entirely negligible.
The antisymmetric sector field variables behave essentially like an
shell of energy to the symmetric sector field variables, which then
collapse to their {\sc gr} form: {\it i.e.}, Schwarzschild-like.
This much is hardly surprising, and indeed, could easily have been
predicted based on a perturbative analysis of the Hamiltonian constraint.
However, what is somewhat more surprising is how quickly the solution to
the Hamiltonian constraint relaxes to its {\sc gr} form outside the
influence of the antisymmetric sector field variables.
In fact, it is clear from 
Figure~\ref{fig:conformal_factor_sample_blow-up} that in the region
$50 \lesssim r \lesssim 60$, $\phi$ has essentially returned to its
{\sc gr} form: the difference between the two curves in the graph
is at its greatest less than $0.5\%$, and this despite the fact that
this region lies within two widths of the central peak of $\psi$.
Indeed, at $r = 50$, $\psi(50) / \psi(r_{\rm peak}) \approx 0.54$,
which is hardly negligible.
This demonstrates the rather surprising result that the antisymmetric 
field variables can very well be non-trivial, and yet the
gravitational system as a whole still behaves very much like {\sc gr}, 
at least as far as the initial-value problem is concerned.

\begin{figure}
\centerline{\epsffile{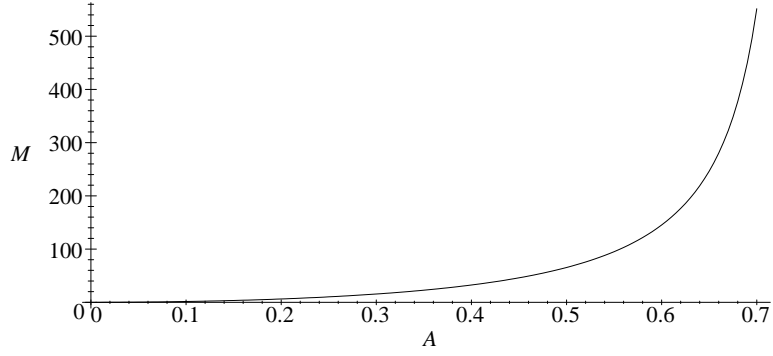}}
\caption{The mass of the gravitational field versus the amplitude 
$A$ of the pulse for massless {\sc ngt},
where $\sigma^2 = 3$ in units where $r_0 = 40$.
The inequalities in Appendix~\protect\ref{sec:Appendix_proof_of_criterion}
give $0.51 \leqslant A \leqslant 0.81$.}
\label{fig:Mass_versus_A_massless}
\end{figure}

\begin{figure}
\centerline{\epsffile{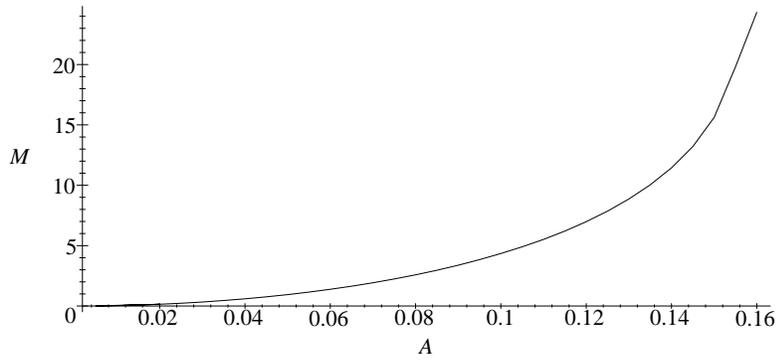}}
\caption{The mass of the gravitational field versus the amplitude 
$A$ of the pulse for massive {\sc ngt},
where $\sigma^2 = 3$ in units where $r_0 = 40$.}
\label{fig:Mass_versus_A_massive}
\end{figure}

\begin{figure}
\centerline{\epsffile{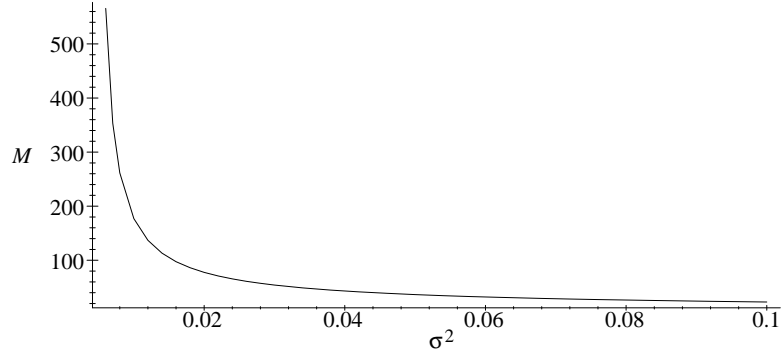}}
\caption{The mass of the gravitational field versus the squared width 
$\sigma^2$ of the pulse for massless {\sc ngt},
where $A = 0.15$, and where $r_0 = 40$ in the same 
units as $\sigma$.
The inequalities in Appendix~\protect\ref{sec:Appendix_proof_of_criterion}
give $0.0046 \leqslant \sigma^2 \leqslant 0.090$ for the
left asymptote, in those same units;
the leftmost point on the graph is at $\sigma^2 \approx 0.0060$, where
$M \approx 566$.}
\label{fig:Mass_versus_sigma_massless}
\end{figure}

\begin{figure}
\centerline{\epsffile{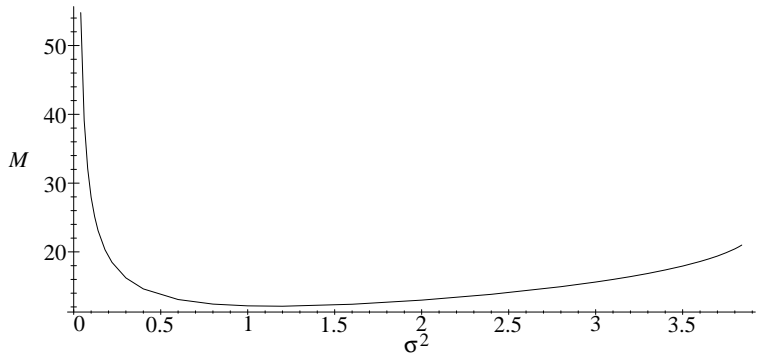}}
\caption{The mass of the gravitational field versus the squared width 
$\sigma^2$ of the pulse for massive {\sc ngt},
where $A = 0.15$, and where $r_0 = 40$ in the same 
units as $\sigma$.}
\label{fig:Mass_versus_sigma_massive}
\end{figure}

\begin{figure}
\centerline{\epsffile{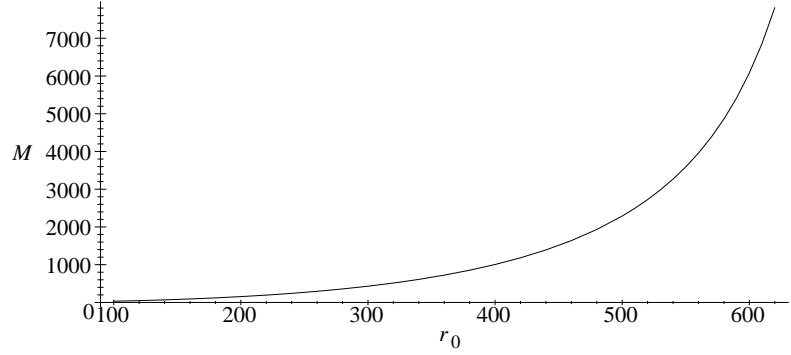}}
\caption{The mass of the gravitational field versus the position 
$r_0$ of the pulse for massless {\sc ngt},
where $A = 0.15$, and where $\sigma^2 = 1.6$ in the
same units as $r_0$.
The inequalities in Appendix~\protect\ref{sec:Appendix_proof_of_criterion}
give $160 \leqslant r_0 \leqslant 750$ in those same units.}
\label{fig:Mass_versus_position_massless}
\end{figure}

\begin{figure}
\centerline{\epsffile{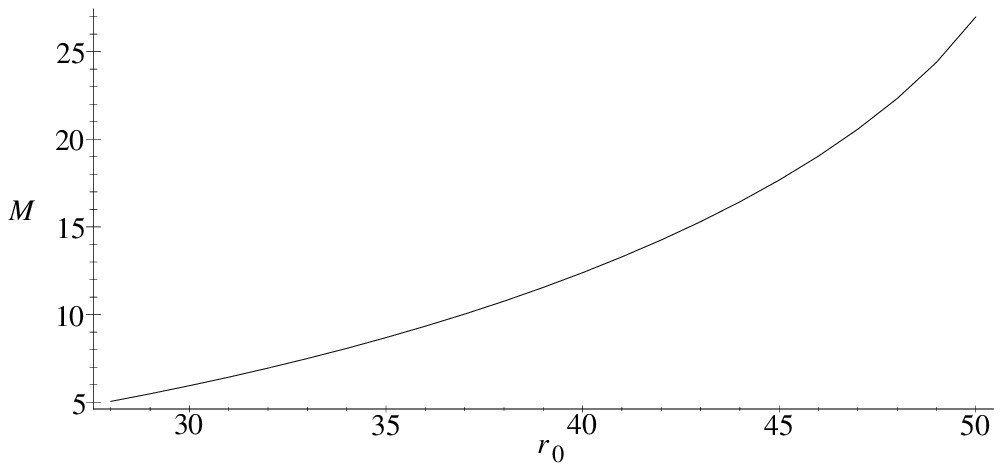}}
\caption{The mass of the gravitational field versus the position 
$r_0$ of the pulse for massive {\sc ngt},
where $A = 0.15$, and where $\sigma^2 = 1.6$ in the
same units as $r_0$.}
\label{fig:Mass_versus_position_massive}
\end{figure}

In Figures~\ref{fig:Mass_versus_A_massless}
through~\ref{fig:Mass_versus_position_massive}
we present plots of the measured
mass of the gravitational field
versus the various parameters of the pulse;
the value of the parameters that remain fixed for the plot are 
given in the figure caption.
For the massless {\sc ngt} calculations, we also supply in the figure
caption the bounds on the parameters determined 
from~(\ref{eq:strong_condition_on_phi})
and~(\ref{eq:weak_condition_on_phi});
note that our initial data allows us to use
the flat-space Sobolev constant.
Of particular importance is the behaviour of the mass as a function of
the pulse width, $\sigma$, for the case of massive {\sc ngt}.
We see from Figure~\ref{fig:Mass_versus_sigma_massive} that there is
a {\em maximum} value of $\sigma$ above which the Hamiltonian constraint
is no longer solvable.
Loosely speaking, this implies that the
pulse~(\ref{eq:not_so_Gaussian_pulse}) must be sufficiently small {\em and}
sufficiently peaked (for a given amplitude)
in order for a solution to the Hamiltonian constraint to exist.
This seems somewhat counter-intuitive; however, when $\mu \ne 0$, 
we see from~(\ref{eq:Hamiltonian_constraint_both_forms})
that the term proportional to $\mu^2$ contains a factor of
$\sin^2\psi$, and is also proportional to $\phi^6$.
An increasing value of the width causes the factor of 
$\sin^2\psi$ to become non-trivial in a larger region of the spatial
slice, thereby increasing the energy contained in the gravitational system.
Consequently, this forces $\phi$ to stray from its assumed 
$\phi \rightarrow 1$ fall-off, and the Hamiltonian constraint
becomes unsolvable.
It therefore
stands to reason that this highly non-linear term can affect the
resulting solution to a great extent.
Of course, a similar behaviour can be observed in the abscence of
this term: loosely
speaking, increasing the width increases the amount of antisymmetric
field that is present on the spatial slice, therefore adding to the
energy of the system.
However, we have found that this occurs at width values on the order of
$\sigma \approx 217$, where $M \approx 3190$.
The inequalities in Appendix~\ref{sec:Appendix_proof_of_criterion} give
$51000 \leqslant \sigma^2 \leqslant 55000$.
We conclude that the non-linear term 
in~(\ref{eq:Hamiltonian_constraint_both_forms})
simply compounds this problem.

We note from the figures that there are clearly defined asymptotes 
in the mass as a function of the parameter being investigated; these
asymptotes correspond to the divergence of the mass as we approach a
region of unsolvability of the Hamiltonian constraint in the parameter
space.
In particular, the mass plots for massless {\sc ngt} have
asymptotes which respect the bounds set by the results of
Appendix~\ref{sec:Appendix_proof_of_criterion}.
For instance, from Figure~\ref{fig:Mass_versus_A_massless}
we see that the asymptote of the mass as a function of the
amplitude lies at $A\approx 0.70$.
This compares well with the bounds of~$A \geqslant 0.51$ and~$A \leqslant 0.81$ set 
by~(\ref{eq:weak_condition_on_phi}) 
and~(\ref{eq:strong_condition_on_phi}), respectively.
Similarly, the asymptote of the mass as a function of the
width squared (see Figure~\ref{fig:Mass_versus_sigma_massless})
lies at about $\sigma^2\approx 0.005$ for
$r_0 = 40$.
This is to be compared with the bounds of~$\sigma^2 \geqslant 0.0046$ 
and~$\sigma^2 \leqslant 0.090$
obtained from~(\ref{eq:strong_condition_on_phi})
and~(\ref{eq:weak_condition_on_phi}), respectively.
Finally, the asymptote of the mass as a function of the position of 
the pulse $r_0$ (see Figure~\ref{fig:Mass_versus_position_massless})
lies at $r_0 \approx 620$ for
$\sigma^2 = 1.6$, in comparison to the values 
of~$r_0 \geqslant 160$ and~$r_0 \leqslant 750$ 
obtained from~(\ref{eq:weak_condition_on_phi})
and~(\ref{eq:strong_condition_on_phi}), respectively.

Generically, the asymptotes of the mass are more clearly defined for 
massless {\sc ngt} than for massive {\sc ngt}.
This is most pronounced when plotting the mass $M$ versus the position
of the pulse $r_0$:
it is clear from Figure~\ref{fig:Mass_versus_position_massive}
that the asymptotic value of $r_0$ has not been reached.
The reason for this is the same as mentioned above:
the non-linear term in~(\ref{eq:Hamiltonian_constraint_both_forms})
begins to grow significantly, and the numerical code fails to find a
solution.
It is also interesting to note that when determining the bounds on the
region of solvability for variations in 
$\sigma$ (see
Figure~\ref{fig:Mass_versus_sigma_massless}),~(\ref{eq:weak_condition_on_phi}) 
seems to give particularly bad results, being off by a 
full order of magnitude.
Although the predicted bound does not contradict the actual result,
and no claim was made as to the accuracy of~(\ref{eq:weak_condition_on_phi}),
the reason for why there is such a large discrepancy is unknown at this time.

\begin{figure}
\centerline{\epsffile{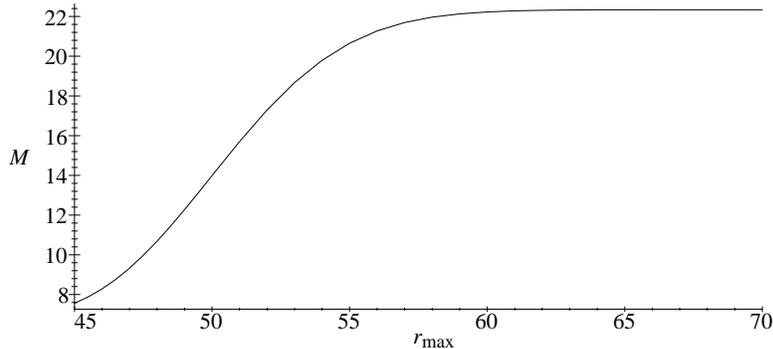}}
\caption{The mass of the gravitational field of massles {\sc ngt}
as a function of the 
outer grid point, $r_{\rm max}$, for the initial conditions
$\gamma^{11} = 1$, $\psi = A[r/r_0]^2 e^{-(r-r_0)^2/\sigma^2}$, 
and $j^2{}_3 = 0$, where
$A = 0.7$, and with $\sigma = 10$ in 
units where $r_0 = 40$.}
\label{fig:Mass_convergence}
\end{figure}

When calculating the {\sc adm} mass of a system, we must take into account
the fact that our numerical grid is not of infinite size, 
whereas~(\ref{eq:ADM_mass_in_NGT_in_general}) implicitly assumes that
the evaluation is performed at infinity.
In Figure~\ref{fig:Mass_convergence} we show how the {\sc adm} mass
varies as the size of the grid is increased, for the initial
data $\gamma^{11} = 1$ and $j^2{}_3 = 0$, with $\psi$ given 
by~(\ref{eq:not_so_Gaussian_pulse}), where
$A = 0.7$, $r_0 = 40$ and $\sigma = 10$.
As expected, the mass quickly converges to a definite value once the
grid is sufficiently large.
In our case, the mass has converged to within $0.01\%$ when
$r_{\rm max} = 70$.
At this point,
$\psi$ has dropped to approximately $0.04\%$ of its
peak value.
This justifies our choice of $r_{\rm max} = r_{\rm peak} + 10\sigma$.

Finally, it is important to verify that the solution $\phi$ obtained 
by numerical means actually satisfies the Hamiltonian constraint, and if
so, how well.
This is somewhat of a thorny issue since, for a 
vacuum spacetime, the correct solution must make
the Hamiltonian constraint vanish.
It is therefore
inheritly difficult to judge how well the solution
satisfies this criterion 
without imposing some arbitrary precision level.
One possibility, similar to that discussed by
Bernstein {\it et.\ al.}
(see Appendix~A.3 in~\cite{bib:Bernstein et al Phys Rev D 50 3760 1994})
is to use the fact that in a non-vacuum spacetime, the Hamiltonian
must be equated to the matter contributions:
${\cal H} = 2 p \rho$.
Thus, the error in the determination of $\phi$ can be viewed as a 
``matter contribution'' to the right-hand side of the Hamiltonian constraint.
Consequently, the mass of this contribution can be calculated and then
compared with the known mass of the hypersurface, which then gives a
measure of the precision of the solution.

\narrowtext
\begin{table}
\caption{The residual Hamiltonian $H_{\rm e}$ 
and the {\sc adm} mass $M$ as a 
function of the grid spacing $\Delta r$ for massless {\sc ngt},
for the initial
data $\gamma^{11} = 1$ and $j^2{}_3 = 0$, 
with $\psi = A[r/r_0]^2 e^{-(r-r_0)^2/\sigma^2}$,
where $A = 0.7$, and $r_0 = 40$ in units where $\sigma = 10$.
\label{tab:convergence_of_constraint_solution}}
\begin{tabular}{ccc}
$\Delta r$ & $M$ & $H_e$ \\
\tableline
0.300 & 22.317 & 4.525$\times 10^{-2}$ \\
0.150 & 22.334 & 1.129$\times 10^{-2}$ \\
0.075 & 22.338 & 2.815$\times 10^{-3}$ \\
0.030 & 22.340 & 4.554$\times 10^{-4}$ \\
0.015 & 22.340 & 1.356$\times 10^{-4}$ \\
\end{tabular}
\end{table}
\widetext

In order to verify that the solution satisfies the Hamiltonian constraint 
to a sufficient degree, we calculate the integral of absolute value of
the Hamiltonian constraint:
\[
H_{\rm e} = \int_\Sigma |{\cal H}| \, d^3 x ,
\]
where the integration is performed over the volume of the slice $\Sigma$.
We refer to $H_e$ as the {\em residual Hamiltonian}, as it is a measure 
of that part of numerically-determined solution
that violates the constraint.
The residual Hamiltonian is then 
compared with the {\sc adm} mass $M$ of the hypersurface.
This process is repeated for successively smaller values of the grid
spacing, until the ratio $H_{\rm e}/M$
is reduced to the desired level of precision.
In Table~\ref{tab:convergence_of_constraint_solution} we list the
results of such a calculation in the case of massless {\sc ngt}
for the initial data
$\gamma^{11} = 1$ and
$j^2{}_3 = 0$, with $\psi$ given 
by~(\ref{eq:not_so_Gaussian_pulse})
where $A = 0.7$, $r_0 = 40$, and $\sigma = 10$.
The {\sc adm} mass and the residual Hamiltonian are given in the same
units as $\Delta r$, the grid spacing.
We see that as the grid spacing is reduced, the {\sc adm} mass converges
to a definite value, in this case $M \approx 22.340$, 
and residual Hamiltonian decreases roughly as $(\Delta r)^2$.
This is in keeping with the fact that
our discretization of the Hamiltonian constraint
was valid to second-order in the grid spacing.

\section{Conclusions}

We have described herein the foundation necessary for performing a
thorough numerical study of spherically symmetric systems in {\sc ngt}.
By virtue of its construction, this formalism is not limited to a static
gravitational field; this is particularly relevant for the 
spherically symmetric, Wyman sector initial data
we have described in this work.
In fact,
the antisymmetric field variables are true dynamical degrees of freedom,
not limited to being diffeomorphically related to a physically-equivalent
static system by some analogue of Birkhoff's theorem.

We have demonstrated that the initial-value problem of {\sc ngt} can
be solved by numerical means, and that the solutions behave essentially
as one would expect, based on a careful physical analysis of the 
situation, as described at the beginning of~\S\ref{sec:Numerical_NGT}.
However, for a given initial data set, we cannot always solve the 
initial-value problem.
We were able to place limits on the solvability of the Hamiltonian constraint
for a particular distribution of the antisymmetric field variables.
In itself, it is not particularly surprising that the initial-value problem
cannot always be satisfied for all possible field configurations.
What is surprising, however, is that certain seemingly acceptable 
configurations do not form a well-posed initial-value problem.
In particular, for the case of the massive theory ($\mu\ne 0$), it was
found that the antisymmetric field variables had to have particularly
rapid fall-off in the asymptotic region in order to guarantee that the
initial slice be asymptotically flat;
otherwise, the final term
in~(\ref{eq:Hamiltonian_constraint}), proportional to both
$\sin\psi$ and $\phi^6$, would feed back into the solution and destroy
the desired asymptotic behaviour.

The behaviour of the {\sc adm} mass, or rather its {\sc ngt} equivalent,
was found to be a good indication of the solvability of the Hamiltonian
constraint.
Indeed, far from regions of unsolvability, the {\sc adm} mass was very
small, corresponding to a weak field r\'egime.
On the other hand, near regions of unsolvability, the {\sc adm} mass 
became very large.
But in the latter cases, the mass always began to display signs of reaching
an asymptote.
Again, this corresponds well with the underlying physical picture: a large
mass cannot be obtained from an asymptotically flat spacetime, so
demanding that the mass be large by increasing the size of the 
antisymmetric field variables causes us to step off the constraint
surface, thereby violating the Hamiltonian constraint.

Of course, there remains much to be explored.
For instance, we have not touched upon the topic of the Cauchy, or
evolution problem in {\sc ngt}.
Once we are satisfied that a given field configuration properly
satisfies the initial-value problem, we can then use the {\sc ngt} 
evolution equations to determine the evolution of the system.
The definition and existence of apparent horizons are also 
pivotal to the structure of {\sc ngt}; this will be dealt with in a
future publication.

Beyond this, we could modify the formalism (and hence the numerical
code) to accommodate matter sources, thereby moving beyond the purely vacuum
case that we have been considering here.
Ideally, matter fields would be combined with an evolutionary approach, and 
we could make some headway into determining precisely
what it is that happens in {\sc ngt} when a stellar object expends 
its fuel and begins to collapse.

Perhaps the most critical aspect of {\sc ngt} is that it remains as a 
candidate for a non-singular theory of gravitation.
These claims are based on the study of a single, spherically symmetric,
vacuum solution 
to the field equations of massless {\sc ngt}
(see~\cite{bib:Cornish and Moffat Phys Lett B 336 337 1994}
and~\cite{bib:Cornish Mod Phys Lett A 9 3629 1994}) as well as a certain
amount of indirect theoretical evidence 
(see~\cite{bib:Moffat J Math Phys 36 5897 1995}).
Meanwhile, some workers have claimed that {\sc ngt} does exhibit the
same singularity properties displayed in {\sc gr}
based on perturbative analyses of the field equations
(see~\cite{bib:Burko and Ori Phys Rev Lett 75 2455 1995}).
Ideally, 
we would prefer to consider initial data that describes 
a stellar model, essentially along 
the lines of the classic Oppenheimer-Snyder collapse problem in
{\sc gr} (see~\cite{bib:Oppenheimer and Snyder Phys Rev 56 455 1939}),
and ``let the thing go''.
However, the mathematical difficulties involved in performing such
feats has for some time relegated the topic to the area of pure
speculation.
Given the formalism we have outlined herein, this ideal is much more 
attainable than it once was.

\acknowledgments

We would like to thank E.\ Seidel, and especially
K.\ Camarda, for their help
during the initial stages of this work,
and N.\ \'O Murchadha for his help with some important theoretical
details.
We would also like to thank J.\ W.\ Moffat and P.\ Savaria for
their many stimulating discussions.
This work was partially
funded by the Natural Sciences and Engineering
Research Council of Canada.
J.\ L\'egar\'e would like to thank the Walter C.\ Sumner Foundation
for their funding of this research.

\appendix

\section{The Spherically Symmetric Field Equations
of NGT in First-Order Form}
\label{sec:Appendix_First-Order_form}

A detailed description of the Hamiltonian form of {\sc ngt} can be found
in~\cite{bib:Clayton thesis}.
In this appendix, we give a brief overview of
the reduction of these field equations for spherically symmetric systems.
In particular, we will be considering exclusively the ``Wyman sector'' of
{\sc ngt}, also known as the electric sector 
(see~\cite{bib:Wyman Can J Math 2 427 1950,%
bib:Bonnor Proc Roy Soc 209 353 1951,%
bib:Vanstone Can J Math 14 568 1962,%
bib:Cornish and Moffat Phys Lett B 336 337 1994}).
The inverse fundamental tensor will therefore take the form
$g^{-1} = e_\bot \otimes e_\bot
- \gamma^{ab} e_a \otimes e_b$,
where the basis vectors are defined by
$N e_\bot = \partial_t - N^a\partial_a$ and
$e_a = \partial_a$

The spherically symmetric form of a nonsymmetric tensor quantity
can be found by a Killing vector analysis
(see~\cite{bib:Clayton thesis}, p.~94).
The spatial part of the inverse fundamental tensor and the
extrinsic curvature are
\[
\gamma^{ab}
\rightarrow \left[
\begin{array}{ccc}
\gamma^{11} & 0 & 0 \\
0 & \gamma^{22} & \gamma^{[23]}/\sin\theta \\
0 & -\gamma^{[23]}/\sin\theta & \gamma^{22}/\sin^2\theta \\
\end{array}\right] 
\qquad\mbox{\rm and}\qquad
K_{ab}
\rightarrow \left[
\begin{array}{ccc}
K_{11} & 0 & 0 \\
0 & K_{22} & j_{[23]}\sin\theta \\
0 & -j_{[23]}\sin\theta & K_{22}\sin^2\theta \\
\end{array}\right] ;
\]
notice how we reserve the indexed symbols for the tensorial
quantity with factors of $\sin\theta$ removed.
The volume element is written 
\[
p\sin\theta = \sqrt{\gamma} = \sqrt{\det(|\gamma_{ab}|)}
= (\gamma^{11}[(\gamma^{22})^2 + (\gamma^{[23]})^2])^{-1/2} 
\sin\theta .
\]
The lapse function depends on the radial variable (and the time), 
$N = N(r, t)$,
while the shift vector reduces to a single non-trivial component:
$|N^a| \rightarrow [N^1, 0, 0]$.
The vectors $|p^a|$ and $|\overline W_a|$ vanish by definition, being both  
native to the ``Papapetrou'' or magnetic sector of {\sc ngt}.
The non-zero components of the fundamental tensor are then
\[
G_{11} = (\gamma^{11})^{-1} , \qquad
G_{22} = \frac{\gamma^{22}}{(\gamma^{22})^2 + (\gamma^{[23]})^2} , \qquad
\mbox{\rm and}\quad
G_{[23]} = -\frac{\gamma^{[23]}}{(\gamma^{22})^2 + (\gamma^{[23]})^2} .
\]
We define the following:
\[
K^1{}_1 = \gamma^{11} K_{11} ,
\qquad
K^2{}_2 = \gamma^{22} K_{22} + \gamma^{[23]} j_{[23]} ,
\qquad\mbox{\rm and}\quad 
j^2{}_3 = \gamma^{[23]} K_{22} - \gamma^{22} j_{[23]} .
\]

For spherically symmetric Wyman sector initial data, the 
non-zero Lagrange multipliers
and connection coefficients can be solved for, to wit:
$a_1 = \partial_r[\ln N]$, and $\sigma^1 = \gamma^{11} a_1$, 
while
\[
\Gamma^1_{11} 
= \case{1}{2}\partial_r[\ln(p^2[(\gamma^{22})^2 + (\gamma^{[23]})^2])] , 
\qquad
\Gamma^2_{12}
= \case{1}{2}\partial_r[\ln(p^2\gamma^{11}[(\gamma^{22})^2
+ (\gamma^{[23]})^2]^{1/2})] ,
\]
and
\[
\Gamma^1_{22} = -\gamma^{11}[G_{22}\Gamma^2_{12} - G_{[23]}\lambda^2_{13}] . 
\]
The non-trivial
components of the tensor $|\lambda^a_{bc}|$ are
\[
\lambda^2_{13} = \case{1}{2}G_{22}\gamma^{[23]}
(\partial_r[\ln(p\gamma^{[23]}) - \partial_r[\ln(p\gamma^{22})]) ,
\qquad\mbox{\rm and}\quad
\lambda^1_{23} = \gamma^{11}(G_{[23]}\Gamma^2_{12}
+ G_{22}\lambda^2_{13}) .
\]
Finally, the components of the {\sc ngt} surface Ricci tensor are 
found using the connection coefficients and 
the components of $|\lambda^a_{bc}|$ given above:
\[
R^{{\rm NS}(3)}_{11} = -2\partial_r[\Gamma^2_{12}] 
+ 2\Gamma^2_{12}(\Gamma^1_{11} - \Gamma^2_{12}) - 2(\lambda^2_{13})^2 ,
\qquad
R^{{\rm NS}(3)}_{22} = \partial_r[\Gamma^1_{22}] + \Gamma^1_{22}\Gamma^1_{11}
+ 2\lambda^1_{23}\lambda^2_{13} + 1 , 
\]
and
\[
R^{{\rm NS}(3)}_{[23]}
= \partial_r[\lambda^1_{23}] + \Gamma^1_{11}\lambda^1_{23}
- 2\Gamma^1_{22}\lambda^2_{13} .
\]

\section{The Conformally-Transformed Field Equations and Constraints for 
Spherically Symmetric NGT}
\label{sec:Appendix_Conformally-transformed_equations}

In order to approach the initial-value problem in {\sc ngt}, in particular
so as to study the solvability of the Hamiltonian constraint, we take the
same route as in {\sc gr}, which consists of conformally scaling the
field variables by a factor $\phi$, thereby transforming the Hamiltonian
constraint into an equation for this conformal 
factor (see~\cite{bib:Lichnerowicz J Math Pures Appl 23 37 1944}).
The variables $|\gamma_{ab}|$ and $|K_{ab}|$ are chosen to transform
in some particular way, which then determines the transformation rules
for the connection coefficients.
Although this can be done in general in {\sc gr}, the process 
is tantamount to inverting the compatibility condition, which in {\sc ngt}
is not nearly as trivial as in {\sc gr}
(see, for instance,~\cite{bib:Tonnelat J Phys Rad 12 81 1951,%
bib:Tonnelat J Phys Rad 16 21 1955}).

We choose $|\gamma_{ab}|$ and $|K_{ab}|$ to transform as
$\gamma_{ab} \rightarrow \phi^4\gamma_{ab}$ and
$K_{ab} \rightarrow \phi^{-2}K_{ab}$
(see~\cite{bib:York 1989}); from this we determine that
$\gamma^{ab}\rightarrow \phi^{-4}\gamma^{ab}$ and
$G_{ab}\rightarrow \phi^4 G_{ab}$,
while $K^a{}_b \rightarrow \phi^{-6} K^a{}_b$
and $j^a{}_b \rightarrow \phi^{-6} j^a{}_b$.
Note that we choose the transformation rule for the
antisymmetric components of the fundamental tensor 
so that the determinant of the fundamental tensor transforms
homogeneously in $\phi$: $p\rightarrow \phi^6 p$.
The non-trivial Lagrange multipliers are found to transform as
$\sigma^1\rightarrow \phi^{-4}\sigma^1$ and $a_1\rightarrow a_1$, 
while the connection coefficients transform as
\[
\Gamma^1_{11} \rightarrow \Gamma^1_{11} + 2\partial_r[\ln\phi] ,
\qquad \Gamma^2_{12} \rightarrow \Gamma^2_{12} + 2\partial_r[\ln\phi] ,
\qquad\mbox{\rm and}\quad
\Gamma^1_{22} \rightarrow \Gamma^1_{22} 
- 2\gamma^{11}G_{22}\partial_r[\ln\phi] .
\]
Finally, the components of the tensor $|\lambda^a_{bc}|$ transform as
\[
\lambda^2_{13} \rightarrow \lambda^2_{13} 
\qquad\mbox{\rm and}\qquad
\lambda^1_{23} \rightarrow \lambda^1_{23} 
+ 2\gamma^{11}G_{23}\partial_r[\ln\phi] .
\]

Despite its complexity, the transformation rule for the momentum
constraint is relatively straightforward:
\begin{eqnarray*}
{\cal H}_1 \rightarrow {\cal H}_1 
- 8 p [K^1{}_1 + 2 K^2{}_2]\partial_r[\ln\phi] 
&=& -[K^1{}_1 + 2K^2{}_2]\partial_r[p]
- K_{11}\partial_r[p\gamma^{11}] + 2 K_{22}\partial_r[p\gamma^{22}]
+ 2 j_{[23]}\partial_r[p\gamma^{[23]}] \\
& & \qquad\mbox{}
+ 4 p \gamma^{22}\partial_r[K_{22}]
+ 4 p \gamma^{[23]}\partial_r[j_{[23]}]
-8 p [K^1{}_1 + 2 K^2{}_2]\partial_r[\ln\phi] = 0 .
\end{eqnarray*}
Meanwhile, the various terms in the Hamiltonian constraint transform
differently, yielding
\begin{eqnarray*}
{\cal H}
&=& 8p\gamma^{11}\phi\partial_r^2[\phi]
- 8 p\phi\partial_r[\phi][\gamma^{11}\Gamma^1_{11}
+ 2\gamma^{22}\Gamma^1_{22} + 2\gamma^{[23]}\lambda^1_{[23]}]
- p\phi^2 R^{{\rm NS}(3)} \\
& & \qquad\mbox{}
- 2p\phi^{-6}[(K^2{}_2)^2 + 2 K^1{}_1K^2{}_2 - (j^2{}_3)^2]
- \case{1}{4}\mu^2 p \phi^6 \gamma^{[23]} G_{[23]} \\
&=& 8p\phi \Delta_\gamma[\phi] - p\phi^2 R^{{\rm NS}(3)}
- 2p\phi^{-6}[(K^2{}_2)^2 + 2 K^1{}_1K^2{}_2 - (j^2{}_3)^2]
- \case{1}{4}\mu^2 p \phi^6 \gamma^{[23]} G_{[23]} = 0 ,
\end{eqnarray*}
where in the last line, we have used the 
definition~(\ref{eq:derivative_definition}) to write the Hamiltonian
in the familiar Lichnerowicz form.

The evolution equations for the components of the fundamental
tensor can be shown to transform in a somewhat simple manner:
\begin{eqnarray*}
\partial_t[\gamma^{11}]
&=& N^1\partial_r[\gamma^{11}] - 2\gamma^{11}\partial_r[N^1]
- 4N^1\gamma^{11}\partial_r[\ln\phi]
- 2N\phi^{-6}(\gamma^{11})^2K_{11} , \\
\partial_t[\gamma^{22}]
&=& \partial_r[N^1\gamma^{22}] 
- 4N^1\gamma^{22}\partial_r[\ln\phi]
- 2N\phi^{-6}[(\gamma^{22})^2 + (\gamma^{[23]})^2]K_{22} , \\
\noalign{\noindent\rm and}
\partial_t[\gamma^{[23]}]
&=& \partial_r[N^1\gamma^{[23]}] 
- 4N^1\gamma^{[23]}\partial_r[\ln\phi]
- 2N\phi^{-6}[(\gamma^{22})^2 + (\gamma^{[23]})^2] j_{23} .
\end{eqnarray*}
The evolution equations for the components of the extrinsic curvature
would be equally as simple, if not for the complicated transformation
rule of the components of $\bar {\cal Z}_{ab}$; in fact, we find that
\begin{eqnarray*}
\partial_t[K_{11}]
&=& N^1\partial_r[K_{11}] + 2 K_{11}\partial_r[N^1]
- 2N^1 K_{11}\partial_r[\ln\phi] - N\phi^2\bar {\cal Z}_{11}
- \case{1}{4}\mu^2 N\phi^6 G_{11}\gamma^{[23]}G_{[23]} , \\
\partial_t[K_{22}]
&=& N^1\partial_r[K_{22}] - 2N^1 K_{22}\partial_r[\ln\phi]
- N\phi^2\bar {\cal Z}_{22} 
+ \case{1}{4}\mu^2 N\phi^6 G_{22}\gamma^{[23]} G_{[23]} , \\
\noalign{\noindent\rm and}
\partial_t[j_{[23]}]
&=& N^1\partial_r[j_{[23]}]
- 2N^1 j_{[23]}\partial_r[\ln\phi] - N\phi^2\bar {\cal Z}_{[23]}
+ \case{1}{4}\mu^2 N\phi^6 [\gamma^{[23]}(G_{22})^2 - G_{[23]}] 
\end{eqnarray*}
where
\begin{eqnarray*}
\bar {\cal Z}_{11}
&=& -4[\phi\partial_r^2[\phi] - (\partial_r[\phi])^2]
+ 2[\partial_r[\ln N] + 2\Gamma^1_{11} 
- 2 \Gamma^2_{12}]\phi\partial_r[\phi] \\
& & \qquad\mbox{}
+ [R^{{\rm NS}(3)}_{11} 
+ \partial_r[\ln N](\Gamma^1_{11} - \partial_r[\ln N])
- \partial_r^2[\ln N]]\phi^2 
+ [(K^1{}_1 + 2K^2{}_2) - 2 K^1{}_1]K_{11} \phi^{-6} , \\
\bar {\cal Z}_{22} 
&=& -2\gamma^{11} G_{22}[\phi\partial_r^2[\phi] + (\partial_r[\phi])^2]
- 2[G_{22}\partial_r[p\gamma^{11}] - \Gamma^1_{22} 
+ p\gamma^{11}G_{22}\partial_r[\ln N]] \phi\partial_r[\phi] \\
& & \qquad\mbox{}
+ [R^{{\rm NS}(3)}_{22} + \partial_r[\ln N]\Gamma^1_{22}]\phi^2 
+ [K^1{}_1K_{22} + 2j_{23} j^2{}_3]\phi^{-6} , \\
\noalign{\noindent\rm and}
\bar {\cal Z}_{[23]}
&=& 2\gamma^{11}G_{[23]}[\phi\partial_r^2[\phi] + (\partial_r[\phi])^2]
+ [2G_{[23]}\partial_r[p\gamma^{11}] + 2\lambda^1_{[23]}
+ 2p\gamma^{11} G_{[23]}\partial_r[\ln N]]\phi\partial_r[\phi] \\
& & \qquad\mbox{}
+ [R^{{\rm NS}(3)}_{[23]} + \partial_r[\ln N]\lambda^1_{[23]}]\phi^2 
+ [K^1{}_1 j_{[23]} - 2K_{22} j^2{}_3]\phi^{-6} .
\end{eqnarray*}

\section{Two Consequences of the Solvability of the Hamiltonian Constraint
for Massless NGT in Spherical Symmetry for a Moment of Time-Symmetry}
\label{sec:Appendix_proof_of_criterion}

The Hamiltonian constraint of 
massless {\sc ngt} for a moment of time-symmetry can be written in the form
$8\phi\Delta_\gamma[\phi] 
= \phi^2 R^{{\rm NS}(3)}$, where the derivative operator
$\Delta_\gamma[\,]$ is defined by
\begin{equation}
\label{eq:derivative_definition}
\Delta_\gamma[\phi] = \gamma^{11}\partial_r^2[\phi]
- \partial_r[\phi](\gamma^{11}\Gamma^1_{11}
+ 2\gamma^{22}\Gamma^1_{22} + 2\gamma^{[23]}\lambda^1_{[23]}) .
\end{equation}
We derive herein two criteria that are consequences of the existence of a
function $\phi$, solution of the Hamiltonian constraint
of massless {\sc ngt}, having the
appropriate properties ($\phi > 0$ and $\phi - 1 \rightarrow 0$ 
as $r^{-1}$ in the asymptotic 
region in order to guarantee finiteness of the 
energy.)\footnote{This proof is a slightly modified 
version of that which appears 
in~\cite{bib:Cantor and Brill Comp Math 43 317 1981};
the notation is similar, although it has been modified when necessary.
It has been modified to fit your screen.}

Let $f$ be a twice differentiable function with asymptotic fall-off.
Since $\phi > 0$, then $f$ can be written as the product
$f = \phi u$, where $u$ is also twice differentiable and has 
the same fall-off as $f$.
Hence,
\begin{equation}
\label{eq:solvability_intermediate}
(\partial_r[f])^2
= \phi^2(\partial_r[u])^2 
+ 2\phi u\partial_r[\phi]\partial_r[u] + u^2(\partial_r[\phi])^2
= \phi^2(\partial_r[u])^2 + \phi\partial_r[\phi]\partial_r[u^2]
+ u^2(\partial_r[\phi])^2 .
\end{equation}
Multiplying both sides by $p\gamma^{11}$ (to form a scalar density)
and integrating gives
\[
\int p\sin\theta\gamma^{11} (\partial_r[f])^2\, d^3x
= \int \sin\theta (p\gamma^{11} \phi^2(\partial_r[u])^2
- \phi u^2\partial_r[p\gamma^{11}]\partial_r[\phi])\, d^3x 
= \int p\sin\theta (\gamma^{11} \phi^2(\partial_r[u])^2
- \phi u^2\Delta_\gamma[\phi])\, d^3x ,
\]
where the cross term $\phi\partial_r[\phi]\partial_r[u^2]$ 
in~(\ref{eq:solvability_intermediate}) has been
integrated by parts;
the resulting surface terms vanish due to the 
fall-off of $u$ and $\phi$.
Now, the first term on the right hand side is manifestly positive, so
that we may conclude that
\[
\int p\sin\theta\gamma^{11}(\partial_r[f])^2\, d^3x
> -\int p\sin\theta u^2\phi \Delta_\gamma[\phi]\,d^3x .
\]
However, assuming that $\phi$ is a solution of the Hamiltonian constraint
and using the fact that $\phi u = f$,
we may write
\begin{equation}
\label{eq:strong_condition_on_phi}
\int p\sin\theta\gamma^{11}(\partial_r[f])^2\, d^3x
> -\frac{1}{8}\int p\sin\theta f^2 R^{{\rm NS}(3)}\,d^3x .
\end{equation}
This condition must be satisfied for all functions $f$ with asymptotic 
fall-off.
Alternately, if the Hamiltonian constraint admits a solution
$\phi$, then the {\sc ngt} scalar curvature
$R^{{\rm NS}(3)}$ is such that all functions $f$ having asymptotic fall-off
will satisfy~(\ref{eq:strong_condition_on_phi}).

For any asymptotically flat Riemannian three-manifold, the 
Sobolev inequality~\cite{bib:Cantor and Brill Comp Math 43 317 1981}
states that there exists a positive constant $C$ 
such that\footnote{Note that our constant $C$ differs
from that of Cantor and~Brill; see, for instance, (10) 
in~\cite{bib:Cantor and Brill Comp Math 43 317 1981}.
In fact, where we use $C$, they use $C^{-1}$.}
\[
\int p\sin\theta\gamma^{11}(\partial_r[\zeta])^2\,d^3x
\geqslant C \Bigl[\int p\sin\theta \zeta^6\, d^3x\Bigr]^{1/3} 
\]
for any infinitely differentiable function $\zeta$ having compact
support: $\zeta \in C_0^\infty$.
For a flat manifold, the Sobolev constant is
$C = \frac{3}{4}2^{2/3} \pi^{4/3}$.
For a non-flat manifold, the value of $C$ is most probably
smaller~\cite{bib:O Murchadha PC}.
Using the Sobolev inequality 
and the H\"older inequality
(see~\cite{bib:Choquet-Bruhat et al 1977}, p.~52),
\[
-\int p\sin\theta \zeta^2 R^{{\rm NS}(3)}\, d^3x 
\leqslant \Bigl[\int p\sin\theta \zeta^6\, d^3x\Bigr]^{1/3} 
\times \Bigl[\int p\sin\theta |R^{{\rm NS}(3)}|^{3/2}\, d^3x\Bigr]^{2/3} ,
\]
we can transform~(\ref{eq:strong_condition_on_phi}) into 
\begin{equation}
\label{eq:weak_condition_on_phi}
\Bigl[\int p\sin\theta 
|R^{{\rm NS}(3)}|^{3/2}\, d^3x\Bigr]^{2/3} \leqslant 8 C .
\end{equation}

The two inequalities~(\ref{eq:strong_condition_on_phi}) 
and~(\ref{eq:weak_condition_on_phi}) give qualitatively different results.
The first of these is used to demonstrate that a data set {\em does not}
support a solution of the Hamiltonian constraint, while the second
is used to prove that a data set {\em does} support a solution of
the Hamiltonian constraint.

\end{document}